\documentclass[aip,jcp,reprint]{revtex4-2}
\usepackage{amsmath,amssymb,bm,graphicx,xcolor,hyperref}

\newcommand{\nn}{\nonumber}
\newcommand{\veps}{\varepsilon}

\newcommand{\ls}{{\rm ls}}

\begin{document}

\title{Flux-induced condensation in a radial geometry}

\author{Yoav Tsori}
\email{tsori@bgu.ac.il}
\affiliation{Department of Chemical Engineering, Ben-Gurion University
of the Negev, Beer-Sheva, Israel}

\date{\today}

\begin{abstract}
Stirring and shear usually suppress phase separation, but transport through a
boundary can induce it. We study a binary mixture that is stable everywhere at
rest and fills a two-dimensional annulus. The inner boundary is semipermeable and
exchanges material with a permeate held at a fixed chemical potential, so that a
steady flux of one component crosses the annulus. Although this driven system carries a steady current and produces entropy,
within Model B dynamics its steady state is an equilibrium profile in an
effective external field that is
logarithmic in the distance from the center. Its strength is not imposed
from outside but set by the permeate and the permeability of the boundary. Above a threshold 
driving flux, a condensed annulus forms
around the inner boundary. It is bounded by a stationary front whose radius
depends exponentially on the ratio of the bulk driving force to the flux. The
front is pinned by the effective field and does not coarsen, and the threshold flux
decreases only as the inverse logarithm of the system size. The driving flux shifts the
phase lines but leaves the critical temperature unchanged. Interfacial tension
gives rise to a critical nucleus and to a fold, at which the condensed state
ceases to exist, with a threshold that we obtain in closed form. Numerical
solutions confirm these predictions and reveal a hysteresis loop bounded by two
saddle-node bifurcations. By symmetry, the theory also
describes cavitation around a sink.
\end{abstract}

\maketitle

%
%
%

\section{Introduction}
\label{sec:intro}

Flow and phase separation have been studied together for several decades, and
in most cases the flow acts against demixing. In simple fluids near a critical
point, shear suppresses composition fluctuations and shifts the coexistence
curve to lower temperatures, so that a sheared mixture can remain homogeneous
below its quiescent critical temperature~\cite{onuki_jpcm1997}. Stirring
arrests the coarsening of domains and, at sufficient amplitude, restores a
homogeneous mixture that would otherwise separate~\cite{thiffeault_pre2007}.
Flow can also induce demixing, as in semidilute polymer solutions and in the
shear banding of complex fluids, but it does so through a coupling of the
stress to the composition or to an internal
structure~\cite{onuki_jpcm1997,olmsted_pre1997,porte_jp2_1997}. In shear
banding, the coexistence of two flowing states shows analogies with
phase transitions induced by external fields in equilibrium, and an effective
nonequilibrium potential has been proposed to account for
them~\cite{porte_jp2_1997}.

Transport through a boundary can have the opposite effect. When a boundary
passes one component of a mixture and retains another, the retained component
accumulates next to it. In membrane filtration, this concentration polarization
raises the solute concentration at the membrane surface above its bulk
value~\cite{song_elimelech_jcsft1995}, and fouling sets in above a critical
permeate flux~\cite{field_jms1995}. The critical flux has been interpreted as a flux-induced transition from a dispersed polarization layer
to a condensed phase at the membrane, and its irreversibility appears as
hysteresis when the flux is raised and then lowered~\cite{bacchin_jms2006}.
Near production wells in gas--condensate reservoirs, a bank of liquid
accumulates where the pressure falls below the dew point, and it impairs
production~\cite{sayed_spe2016}. In drying films and during membrane formation,
the loss of solvent through a free surface drives demixing near that surface.
Either the mixture is quenched into the two-phase region from the outset, so
that the flux shapes the morphology rather than creating it~\cite{bouyer_jms_2021},
or an initially homogeneous film crosses the binodal first at the surface, and
an ordering front then propagates until the whole film has
separated~\cite{clark_epl_2007}. In neither case is there a steady state.

A close analogue exists for heat transport. In liquid $^4$He near the
superfluid transition, a steady heat current leads to a stationary interface
between the normal and the superfluid phases. In a finite system, the onset
of the two-phase state is first order, with a range of
bistability~\cite{onuki_jltp1983}. The interface induced by heat flow has been
compared with the one induced by gravity~\cite{onuki_jpcm1998}, and the heat
current also depresses the transition temperature~\cite{duncan_ahlers_prl1988}.
There, however, the transported quantity is energy rather than the order
parameter, and the underlying transition is continuous.

These examples raise a question that can be posed in the simplest conserved
description of a mixture, Model B~\cite{hohenberg_halperin_rmp1977}, with a
square-gradient free energy~\cite{cahn_hilliard_jcp1958}. Consider a binary
mixture that is stable everywhere at rest, occupies a two-dimensional
annulus, and exchanges material through its inner boundary with a permeate held
at a fixed chemical potential, so that a steady flux of the order parameter
crosses the annulus. Can this flux, with no chemical reaction, no coupling to
the stress, and no applied field, create a stationary phase boundary?
In ordinary phase separation, domains coarsen without limit and no length is
selected~\cite{bray_advphys1994}. A front held at a fixed position is
therefore qualitatively different.

We show that the answer is yes, and that the problem has a familiar structure.
A single integration of the steady-state equation turns the flux into a
spatially varying bias of the chemical potential, which acts as an external
field conjugate to the composition. The problem therefore belongs to the class
of phase transitions in nonuniform external fields, which includes fluids in
gravity~\cite{hohenberg_pra_1972,moldover_rmp1979,dickinson_scott_prl1975},
mixtures in a
centrifuge~\cite{scriven_jcis1986,tsori_crphysique_2007,horinek_pccp_2021,colfen_research2023},
and liquids in electric field
gradients~\cite{ttl_nature2004,tsori_pnas2007,tsori_rmp2009,tsori_jpcb2011,tsori_jcp2021}.
A localized pump is likewise known to influence a critical fluid over long
distances~\cite{kafri_prl2022}. A general mean-field treatment of such problems,
in which the external term in the free energy is the product of a function of
the order parameter and a function of position, was given in
Ref.~\cite{kroll_tsori_jcp2022}, where the present flux-driven case appeared
briefly as an example. Here, however, the field is generated by transport rather
than applied from outside. Its strength is not a control parameter but is fixed
by the chemical potential of the permeate and the permeability of the boundary,
and the steady state carries a current and produces entropy, even though its
composition profile coincides with an equilibrium profile in the effective
field. In a two-dimensional radial geometry the field is, moreover, logarithmic
in the distance, and this marginal form has consequences that we develop below.

The system is thus driven out of equilibrium, in the manner of boundary-driven
diffusive systems, in which reservoirs at different chemical potentials maintain
a steady
current~\cite{schmittmann_zia_physrep1998,eyink_lebowitz_spohn_cmp1990}. For
some such models, the stationary state is a Gibbs measure with a linearly
varying chemical potential~\cite{eyink_lebowitz_spohn_cmp1990}. Below the
critical temperature, boundary-driven systems in planar geometry sustain a
stationary interface between phases imposed by the two reservoirs;
macroscopically, the profile solves a stationary Stefan
problem~\cite{demasi_arma_2011}. The current there can even flow up the density
gradient~\cite{colangeli_jpa2017,colangeli_pre2018}. Here, by contrast, the
mixture is homogeneous and stable at rest, the condensed phase is created by the flux, and the
radial geometry selects a unique front radius.

The resulting picture differs in several respects from ordinary phase
separation. Above a threshold flux, a condensed annulus forms around the inner
boundary, separated from the surrounding mixture by a stationary front whose
radius is selected by the flux. The front is pinned by the field and does not
coarsen. The threshold flux decreases only slowly as the system grows, so that
the transition retains a memory of the system size. The critical temperature is
unchanged by the flux, whereas the curvature of the front gives rise to a
critical nucleus and limits the range over which the condensed state exists.
The condensed and the front-free states both exist over a range of fluxes
bounded by two saddle-node bifurcations, so that the transition is hysteretic.
By symmetry, the same theory describes a dilute region that forms around a sink
in a dense mixture.

The paper is organized as follows. Section~\ref{sec:general} formulates the
model for a partially permeable inner boundary in general, with advection. It then focuses 
on the  case without flow, where the effective field is exactly
logarithmic. Section~\ref{sec:caseA} describes the large-scale
composition profiles and the position of the front (Sec.~\ref{sec_coex}), the
pressure (Sec.~\ref{sec:pressure}), the domain wall and the curvature correction
to the front (Sec.~\ref{sec:gibbs_thomson}), numerical solutions
(Sec.~\ref{sec:numerics}), the shifted binodal lines (Sec.~\ref{sec:shifted_binodal}),
the critical nucleus and the fold (Sec.~\ref{sec:nucleation}), and metastability
and hysteresis (Sec.~\ref{sec:metastab_hyst}). Section~\ref{sec:discussion}
discusses the results and their relation to other systems. The case of
advection through a rejecting boundary, which describes filtration, will be
treated in a separate paper.

\section{General model: radial advection, diffusion, and a partially permeable
inner boundary}
\label{sec:general}

\subsection{Geometry and fields}

We consider a binary mixture or a liquid--vapor system confined within a
two-dimensional annular geometry of depth $w$ in the $z$ direction, bounded by
an inner radius $r=a$ and an outer radius $r=L$. Fluid is withdrawn through the
inner boundary at a fixed volumetric rate $Q$. We use $\psi$ for the
dimensionless volume fraction of the ``heavy'' (condensable) component.
At the outer boundary, the composition is held at a bath
value $\psi(L)=\psi_0$ corresponding to a stable, homogeneous, phase.

For an incompressible mixture, the radial volume-averaged velocity $u_r$ is 
fixed by conservation,
\begin{equation}\label{eq:u_r}
u_r(r)=-\frac{\alpha}{r},\qquad \alpha\equiv\frac{Q}{2\pi w},
\end{equation}
so that $\alpha>0$ corresponds to inward suction. The composition obeys
\begin{eqnarray}
\frac{\partial\psi}{\partial t}+\nabla\cdot\bm{j}&=&0\label{eq:general_dynamics},\\
\bm{j}&=&\psi\bm{u}-M(\psi)\nabla\mu\label{eq:flux},
\end{eqnarray}
where the Onsager transport coefficient (mobility) $M$ is in general
composition dependent. Model B is recovered for $\alpha=0$ and a constant $M$.

The thermodynamics is described by the Ginzburg--Landau functional
\begin{equation}\label{eq:F}
F =\frac{wk_B T}{v_0} \int \left[ f_b(\psi, T)
+ \frac{1}{2}B(\nabla\psi)^2 \right] d^2r ,
\end{equation}
where $v_0$ is a molecular volume, $k_B$ is Boltzmann's constant, and $T$ is the absolute 
temperature. The bulk free energy density $f_b$ is assumed to be written as a
symmetric Landau expansion in $\phi=\psi-\psi_c$, where $\psi_c=1/2$ is the critical composition:
\begin{equation}\label{eq:f_b}
f_b(\phi, T) = f_0 + \frac{1}{2}b\phi^2 + \frac{1}{4}c\phi^4 .
\end{equation}
The parameter $b\sim(T-T_c)$ measures the distance from the critical
temperature $T_c$, and $c>0$ ensures stability. The square-gradient coefficient $B$
sets the surface tension and the domain-wall width. Throughout, $\mu$ denotes
the chemical potential per molecular volume in units of $k_BT/v_0$,
\begin{equation}\label{eq:chempot}
\mu = \frac{\delta f}{\delta \phi} = f_b'(\phi)- B\nabla^2\phi ,
\end{equation}
so that $\mu$, $f_b$, and the pressure below all have the same units, and the
prefactor $wk_BT/v_0$ of Eq.~\eqref{eq:F} never appears explicitly.

We take the temperature to be uniform, so that $b$ is the same everywhere. This
is an approximation. In a steady state, the composition current crosses any
phase boundary continuously, so that heat of mixing is absorbed or released
there, and the current also dissipates energy. The resulting temperature
variations are small when heat conduction is fast compared with mass diffusion.
This is the case in liquid mixtures, where the thermal diffusivity ($\sim 10^{-7}$\,m$^2$/s) typically
exceeds the mass diffusivity ($\sim 10^{-9}$\,m$^2$/s) by about two orders of magnitude.

\subsection{The inner boundary}

The inner boundary is in general \emph{partially permeable}: the withdrawn
stream is depleted in the ``heavy'' component relative to the local mixture. We
encode this in a rejection coefficient $\sigma\in[0,1]$~\cite{staverman_rtc1951}, defined so that the
permeate carries the heavy component at composition $(1-\sigma)\psi_a$, where
$\psi_a\equiv\psi(a)$. The total flux there is 
\begin{equation}\label{eq:sigma_bc}
j_r(a)=(1-\sigma)\,\psi_a\,u_r(a).
\end{equation}
Here, $\sigma=0$ describes a fully transparent boundary that passes the mixture
at its local composition, and $\sigma=1$ a perfectly rejecting boundary that
passes only the light component. The left-hand side of Eq.~\eqref{eq:sigma_bc} is the \emph{total} 
flux, diffusive contribution included; the right-hand side is a constitutive statement about the 
composition of the withdrawn stream.

A real membrane also leaks diffusively. In linear response, this adds a flux
$-\kappa_m[\mu(a)-\mu_p]$, where $\mu(a)$ is the chemical potential at the
wall, $\mu_p$ that of the permeate on the inner side of the membrane, and
$\kappa_m=M_m/\ell_m$ an interfacial transfer coefficient obtained by
homogenizing a layer of thickness $\ell_m$ and mobility $M_m$. The generalized
boundary condition is thus
\begin{equation}\label{eq:sigma_bc_leaky}
j_r(a)=(1-\sigma)\,\psi_a\,u_r(a)-\kappa_m\big[\mu(a)-\mu_p\big].
\end{equation}
This is the Kedem--Katchalsky form of the boundary
condition~\cite{kedem_katchalsky_bba1958}, with the composition on the feed side in 
place of the mean concentration on the two sides of the membrane, and the difference in chemical 
potential in place of the difference in osmotic pressure. Here, $\sigma$ and $\kappa_m$ appear as 
the two independent parameters of the interface.  
The limit $\kappa_m\to\infty$ pins $\mu(a)=\mu_p$, a Dirichlet
condition describing a boundary that has no resistance. The remaining conditions are the natural
(neutral-wall) ones together with the bath value,
\begin{equation}\label{eq:bcs}
\left. \nabla\phi \cdot \bm{n} \right|_{r=a} = 0,
\qquad
\left. \nabla\phi \cdot \bm{n} \right|_{r=L} = 0,
\qquad
\phi(L) = \phi_0 .
\end{equation}
Equations~\eqref{eq:sigma_bc_leaky}--\eqref{eq:bcs} are four conditions for the
fourth-order problem Eq.~\eqref{eq:general_dynamics}.

\subsection{Steady state: the general relation}

In steady state with axisymmetry, $rj_r=\text{const}$, where
the constant is given by Eq.~\eqref{eq:sigma_bc_leaky}. Equations~\eqref{eq:flux}
and \eqref{eq:u_r} then yield
\begin{equation}\label{eq:general_steady}
\frac{\partial\mu}{\partial r}
=\frac{1}{M(\psi)\,r}
\Big\{\alpha\big[(1-\sigma)\psi_a-\psi(r)\big]+a\kappa_m\Delta\mu_a\Big\},
\end{equation}
where we defined $\Delta\mu_a\equiv \mu(a)-\mu_p$. The $1/r$
structure is geometric: it follows from $rj_r=\text{const}$ in two dimensions
and is independent of the constitutive laws. The chemical potential is biased
\emph{logarithmically} in $r$ whenever the brace is constant, and this is the
origin of exponential laws derived below.

The brace contains two physically distinct drivings: advective delivery through
$\alpha$ and boundary leakage through $\kappa_m$. Written in this grouping,
Eq.~\eqref{eq:general_steady} is regular as $\alpha\to0$, and the two mechanisms
behave quite differently. 

In this paper, we treat the case of no flow, $\alpha=0$, in which only
boundary leakage drives the system and the logarithmic field is exact
(Sec.~\ref{sec:caseA}). The opposite case, a perfectly rejecting boundary
without leakage ($\sigma=1$, $\kappa_m=0$) under inward suction ($\alpha>0$),
gives $\partial_r\mu=-(\alpha/M)\,\psi/r$. There, the total flux vanishes, the
diffusive flux exactly cancelling the advective delivery, and the heavy
component accumulates at the inner boundary: this is concentration
polarization. The field is then \emph{not} logarithmic, since the numerator
varies with $r$ through $\psi(r)$; it reduces to the form treated here only in
the weak-polarization limit $\psi\simeq\psi_0$.
The advective problem will be treated elsewhere.

\section{No flow, $\alpha=0$} \label{sec:caseA}

With no volume flow, the inner boundary is a semipermeable window onto a
permeate held at $\mu_p$, and only leakage drives the system. 
We non-dimensionalize the spatial coordinates by scaling all lengths 
relative to the inner radius $a$, i.e., $r'=r/a$, $L'=L/a$, and drop the primes throughout.
For a constant mobility $M$ and in axisymmetric polar coordinates, 
Eq.~\eqref{eq:general_steady} is written as
\begin{eqnarray}
	\veps\left(\phi_{rrr}+\frac{1}{r}\phi_{rr}-\frac{1}{r^2}\phi_r\right)-f_b''(\phi)\phi_r=
	\frac{J_0a}{r},\label{eq:phi_3rd_order}\\
	J_0 a\equiv -\Lambda \Delta\mu_a,\qquad \Lambda\equiv \frac{a\kappa_m}{M}\label{eq:J0_caseA},
\end{eqnarray}
and we have used $\veps\equiv B/a^2$. This equation is subject to the boundary conditions 
$\phi'(1)=\phi'(L)=0$ and $\phi(L)=\phi_0$.

Equation~\eqref{eq:phi_3rd_order} integrates to
\begin{equation}\label{eq:chempot2}
\mu(r)=-J_0 a \ln (r) + C,
\end{equation}
where $C$ is an integration constant. The forced transport thus acts exactly as a spatially 
inhomogeneous external field conjugate to $\phi$. The profiles $\phi(r)$ increase monotonically from
$\phi_0$ as $r$ decreases from $L$ to $1$: the driving shifts the local
chemical potential logarithmically, pushing the system toward liquid--vapor
coexistence near the inner radius. 

What is not known in advance is the value of the drinving $J_0a$. Equation~\eqref{eq:J0_caseA} 
involves $\mu(a)$, which is part of the solution, so $J_0a$
is an eigenvalue determined together with the profile. 

{\bf Closure.} The eigenvalue can be eliminated in closed form to leading
order in $\veps$. Since $\phi'(L)=0$, the only gradient contribution to $\mu$ at
the outer boundary is $-\veps\,\phi''(L)=O(\veps)$. Hence
$\mu(L)\simeq f_b'(\phi_0)$ and $C\simeq f_b'(\phi_0)+J_0a\ln L$, and, since
$\mu(a)=C$, Eq.~\eqref{eq:J0_caseA} yields a relation that is linear in $J_0a$,
\begin{equation}\label{eq:J0_closure}
J_0 a=\frac{\Lambda\big[\mu_p-f_b'(\phi_0)\big]}{1+\Lambda\ln L}
\qquad\big[+O(\veps)\big].
\end{equation}
This $O(\veps)$ error affects the \emph{value} of $J_0a$ and not the form of the
solution, which remains logarithmic. If greater accuracy is wanted,
$\mu(a)$ may be read off from the computed profile and
Eq.~\eqref{eq:J0_caseA} iterated.

There are two limits to Eq.~\eqref{eq:J0_closure}. For an impermeable boundary, 
$\Lambda\to 0$, the limit is $J_0a\to 0$: there is no leak and no driving. In the opposite extreme, 
$\Lambda\to\infty$, one obtains $J_0a\to[\mu_p-f_b'(\phi_0)]/\ln L$, which means
that the mixture at the wall is in equilibrium with the permeate behind the wall. 

Two cases follow:

\begin{itemize}

\item Condensation (liquid injection): $0<\mu(a)<\mu_p$ and $J_0a>0$.

In this case, liquid is injected at $r=a$. The permeate's chemical potential is positive and high; it 
decreases across the membrane and continues to decrease with $r$. The solute diffuses outward 
($j_r>0$). As we show in Sec.~\ref{sec_coex}, for a sufficiently large system, there is a finite distance 
where $\mu=0$, and coexistence with gas occurs.  At larger radii, the fluid is gas. 

\item Cavitation (gas suction): $0>\mu(a)>\mu_p$ and $J_0a<0$.

Here, the ``wellbore'' contains gas at a low density and with a negative $\mu_p$. The chemical 
potential increases with $r$. The flux is inward ($j_r<0$), with material leaving through the membrane 
at $r=a$. The fluid at the outer boundary is liquid. If the system is large enough, coexistence with the
liquid bath is possible at a distance $r$ where $\mu=0$. 

Note that cavitation is obtained from condensation by the operation $\phi\to -\phi$ and $J_0\to -J_0$. 

\end{itemize}

\subsection{Flux-induced coexistence: bulk domains and interface location}\label{sec_coex}

In the case considered here, the logarithmic field, Eq.~\eqref{eq:chempot2}, is
exact, and the mixture is in a spatially inhomogeneous external field. As for
other such fields~\cite{kroll_tsori_jcp2022}, the bulk critical point becomes a
line of second-order phase transitions that terminates at a surface critical
point. Below this point, a line of first-order transitions separates the stable
and unstable regions of the phase diagram. The flux term in
Eq.~\eqref{eq:chempot2} is independent of $\phi$, and therefore the temperature
of the surface critical point equals the bulk critical temperature.

{\bf Bulk domains.} 
Equations~\eqref{eq:chempot} and~\eqref{eq:chempot2} give
\begin{equation}\label{eq:mu_profile}
f_b'(\phi)-\veps\nabla^2\phi=-J_0 a \ln (r) + C .
\end{equation}
Because $B$ scales with molecular dimensions and $a$ is macroscopic, the
gradient parameter $\veps=B/a^2$ is small, and we work in the strongly segregated
limit. The domain then decomposes into large-scale bulk regions separated by
thin boundary layers. Far from any interface, the gradient terms are negligible. Since the outer
boundary condition $\phi'(L)=0$ makes the gradient term vanish there as well, the large-scale 
composition $\phi_\ls$ obeys the algebraic cubic equation
\begin{equation}\label{eq:large_scale}
b\phi_\ls + c\phi_\ls^3=-J_0 a \ln\left(\frac{r}{L}\right)+f'_b(\phi_0).
\end{equation}
%
%
\begin{figure}[htbp]
\centering
 \includegraphics[width=0.48\textwidth]{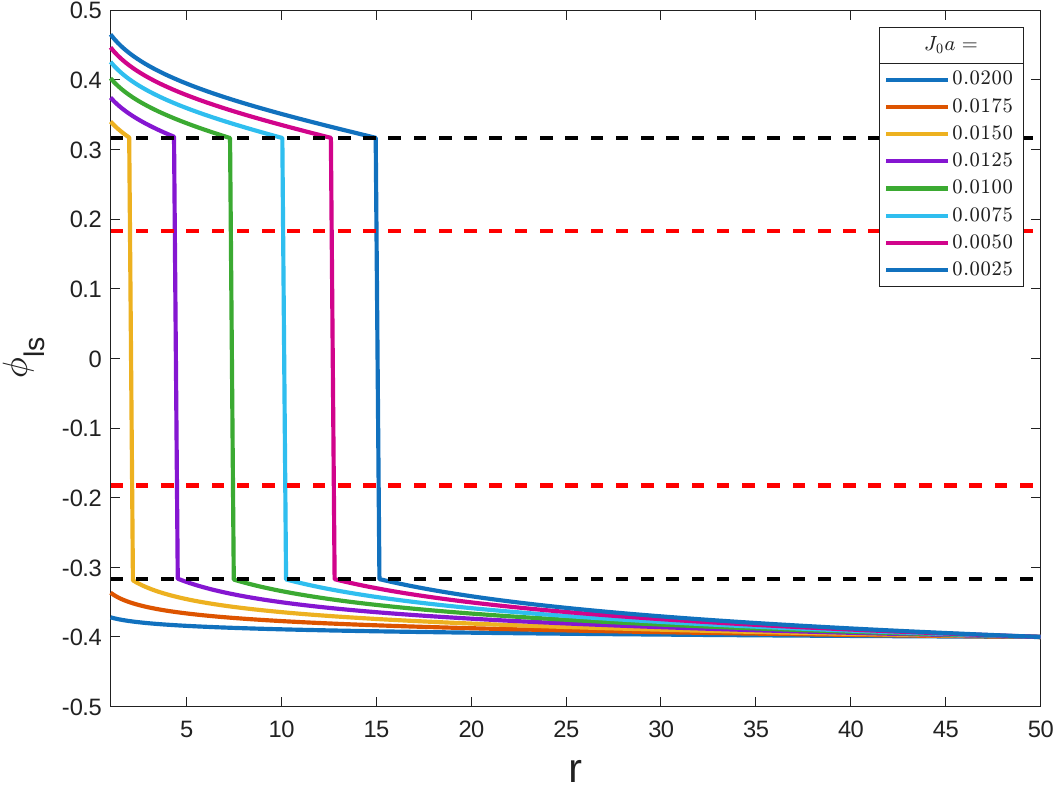}
\caption{Analytical large-scale composition profiles obtained by solving the
cubic equation \eqref{eq:large_scale} for the values of $J_0a$ in the legend.
At large distances, $\phi_\ls\approx\phi_0$, which corresponds to the gas phase. As
$r$ decreases, $\phi_\ls$ increases, and at $r=R$ the profile jumps to the
liquid branch; for $r<R$ the liquid composition increases monotonically. The
horizontal dashed lines are the binodal (black, $f_b'=0$) and spinodal (red,
$f_b''=0$) compositions. The two smallest fluxes lie below the threshold of
Eq.~\eqref{eq:J0a_min} and show no transition.
We used $\phi_0=-0.4$, $b=-0.1$, $c=1$, $L=50$, and $\veps=0.05$.}
\label{fig_phi_ls_profiles_var_J0}
\end{figure}
Equation~\eqref{eq:large_scale} states that the composition adapts locally and
monotonically to the logarithmic field. At low enough temperature (sufficiently
negative $b$), the argument enters the coexistence region, and the cubic equation for $\phi_{\rm ls}$ 
has three real roots. The outer two are the locally stable liquid ($\phi_\ls^l$) and gas
($\phi_\ls^g$) branches; the middle root is unstable. The system follows the branch whose chemical potential lies on the
appropriate side of the coexistence value. For the symmetric free energy
\eqref{eq:f_b} and for shifted chemical potentials that do not depend on $\phi$, the coexistence chemical potential is $\mu_{\rm coex}=0$, so the liquid branch is selected where $\mu(r)>0$ and the gas branch where $\mu(r)<0$.
Figure~\ref{fig_phi_ls_profiles_var_J0} shows $\phi_\ls(r)$ for increasing
$J_0a$. For the two smallest fluxes, the profile remains on the gas branch
throughout; above the threshold derived below, it jumps to the liquid branch at
a radius $R$ that moves outward as $J_0a$ grows.

\begin{figure}[htbp]
\centering
 \includegraphics[width=0.48\textwidth]{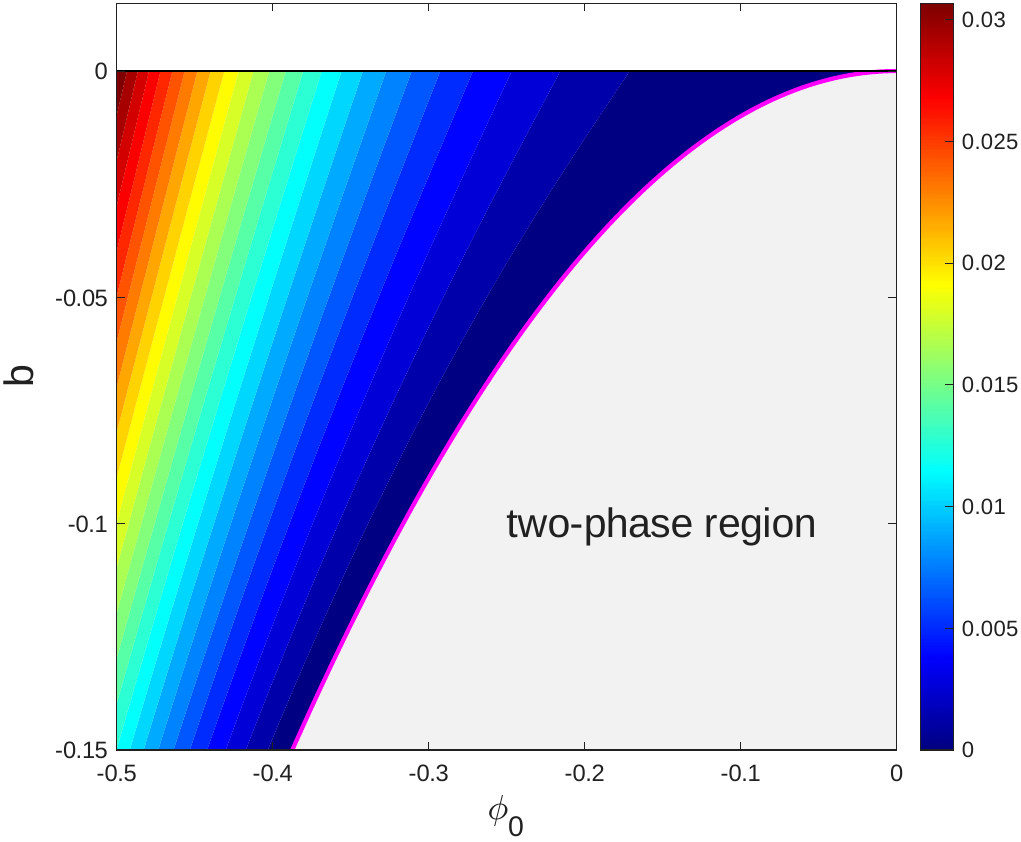}
\caption{The minimal flux $(J_0a)_{\rm min}$ for condensation at $r=a$,
Eq.~\eqref{eq:J0a_min}, in the plane of vapor composition and temperature
$(\phi_0,b)$. The critical point is at $(\phi,b)=(0,0)$; the binodal is magenta,
and the two-phase region, where the mixture separates without any driving, is
shaded. Parameter values are the same as in Fig. \ref{fig_phi_ls_profiles_var_J0}.}
\label{fig_J0a_min}
\end{figure}

The first-order transition therefore occurs at the radius $R$ where the shifted
chemical potential obeys $\mu(R)=0$, 
\begin{equation}\label{eq:R_criterion}
-J_0 a \ln\!\left(\frac{R}{L}\right) + f_b'(\phi_0)=0 .
\end{equation}
In the bulk gas phase, $f_b'(\phi_0)<0$, and hence $R<L$. The solution for $R$ is
\begin{equation}\label{eq:radius_analytical}
R = L \exp\left(\frac{f_b'(\phi_0)}{J_0 a} \right).
\end{equation}
The front radius depends \emph{exponentially} on the ratio of the bulk
thermodynamic chemical potential to the imposed flux. Since $R>a$ is required for a
condensed region to exist at all, the minimal flux for condensation is
\begin{equation}\label{eq:J0a_min}
(J_0a)_{\rm min}=-\frac{f_b'(\phi_0)}{\ln L}.
\end{equation}
The threshold vanishes only logarithmically as $L\to\infty$: the transition is
not a local bulk property but retains a memory of the system size and is a 
consequence of the marginal, logarithmic form of the effective potential
in two dimensions. Figure~\ref{fig_J0a_min} shows this minimal flux in the 
composition--temperature plane $(\phi_0,b)$. Close to the bulk binodal line, $(J_0a)_{\rm min}$ 
is small, and it increases with the distance from it.

We emphasize that the phase separation here is pinned at a finite radius $R$ and, unlike
classical phase separation, does not coarsen. This is a consequence of the dependence of the effective 
driving on $r$.

\subsection{Pressure profiles and mechanical balance}
\label{sec:pressure}

The pressure tensor associated with the functional of Eq.~\eqref{eq:F} is
\begin{equation}\label{eq:ptensor}
\bm{p}=\left[ p_b-\veps\,\psi\nabla^2\psi
-\tfrac12\veps|\nabla\psi|^2 \right]\mathbf{I}
+\veps\,(\nabla\psi \otimes \nabla\psi),
\end{equation}
where the bulk pressure $p_b$ is
\begin{equation}
	p_b=\psi f_b'(\psi) - f_b(\psi).
\end{equation}
With Eq.~\eqref{eq:f_b}, $p_b$ is expressed in terms of $\phi$ as
\begin{equation}\label{eq:p_b}
p_b(\phi)=\tfrac12 b\phi(1+\phi)+\tfrac12 c\phi^3\left(1+\tfrac{3}{2}\phi\right)
-f_0 .
\end{equation}
In azimuthal symmetry, the $\theta$ derivatives vanish, and the pressure 
tensor has two components: $p_{rr}$ and $p_{\theta\theta}$, expressed using $\phi$ as 
\begin{eqnarray}\label{eq:prr}
p_{rr}&=&p_b-\veps\left(\phi+\tfrac12\right)\left(
\frac{\partial^2\phi}{\partial r^2}+\frac{1}{r}\frac{\partial\phi}{\partial r}
\right) + \tfrac{1}{2}\veps\left(\frac{\partial\phi}{\partial r}\right)^2,\nn\\
p_{\theta\theta}&=&p_{rr}-\veps\left(\frac{\partial\phi}{\partial r}\right)^2.
\end{eqnarray}
The pressure is spatially nonuniform due to the axial symmetry and the surface-tension
forces (the terms proportional to $\veps$). The 
drive by the external flux contributes to the 
pressure implicitly through the boundary conditions on $\phi$. 

Unlike the equilibrium case, the stress tensor here is \emph{not} divergence
free. We use $\nabla\cdot\bm{p}=-\psi\nabla\mu$ together with
Eq.~\eqref{eq:mu_profile} and obtain
\begin{equation}\label{eq:mech_balance}
\frac{dp_{rr}}{dr}+\frac{p_{rr}-p_{\theta\theta}}{r}
=\psi\,\frac{J_0a}{r} .
\end{equation}
The driving therefore exerts a distributed body force on the fluid, largest
where the composition is largest, i.e., in the condensed region.

\subsection{Domain wall, surface tension, and curvature correction}
\label{sec:gibbs_thomson}

Near $r=R$, the composition varies over a
distance much shorter than $R$, so the curvature term $\phi'/r$ in
Eq.~\eqref{eq:mu_profile} is smaller than $\phi''$ by a factor
$O(\sqrt{\veps}/R)$ and may be dropped at leading order. Since
$\mu(R)=\mu_{\rm coex}=0$, Eq.~\eqref{eq:mu_profile} reduces locally to the
planar problem $\veps\,\phi''=f_b'(\phi)=b\phi+c\phi^3$, whose solution joining
the liquid branch at $r<R$ to the gas branch at $r>R$ is
\begin{eqnarray}\label{eq:domain_wall}
\phi_{\rm dw}(r)&=&-\phi_b\tanh\big[\kappa\,(r-R)\big],\\
\phi_b&=&\sqrt{-b/c} ,\qquad
\kappa=\sqrt{\frac{-b}{2\veps}} .\nn
\end{eqnarray}
The wall width is $\kappa^{-1}=\sqrt{2\veps/(-b)}$. Far from the wall, the
profile relaxes onto the branch values exponentially, at the rate
$2\kappa=\sqrt{f_b''(\pm\phi_b)/\veps}$ set by the curvature of $f_b$ on the
branches, $f_b''(\pm\phi_b)=-2b>0$.

Equation~\eqref{eq:R_criterion} treats the interface as flat. To use
the Gibbs--Thomson shift of the chemical potential, $\gamma/(\Delta\phi\,R)$, with
the composition jump $\Delta\phi=\phi_\ls^l-\phi_\ls^g=2\phi_b$, we need to calculate the surface tension using the stress 
components introduced above, 
\begin{eqnarray}\label{eq:gamma_def}
	\gamma&=&\int\Big[\big(p_{rr}-p_{\theta\theta}\big)
	-\big(p_{rr}-p_{\theta\theta}\big)_\ls\Big]\,dr\\
	&=&\veps\int\left[\left(\frac{\partial\phi}{\partial r}\right)^{2}
	-\left(\frac{\partial\phi_\ls}{\partial r}\right)^{2}\right]dr .\nn
\end{eqnarray}
As a crude approximation valid for a cylindrical liquid region of sufficiently large $R$, we 
use the surface tension of the planar wall, Eq.~\eqref{eq:domain_wall},
which yields
\begin{equation}\label{eq:gamma_tanh}
\gamma=\frac{4}{3}\veps\phi_b^2\kappa=\frac{2\sqrt{2}}{3}\frac{\sqrt{\veps}\,(-b)^{3/2}}{c}.
\end{equation}
Since lengths are measured in units of $a$, Eq.~\eqref{eq:gamma_tanh} gives
$\gamma$ in units of $(k_BT/v_0)\,a$. In physical units,
$\gamma=(k_BT/v_0)\,(2\sqrt2/3)\sqrt{B}\,(-b)^{3/2}/c$, which is independent of
$a$: the apparent dependence on $a$ through $\veps=B/a^2$ is only that of the
length unit.

The corrected criterion replacing Eq.~\eqref{eq:R_criterion} is
\begin{equation}\label{eq:R_GT}
	\mathcal{F}(R)\equiv
	-J_0 a \ln\!\left(\frac{R}{L}\right) + f_b'(\phi_0)
	-\frac{\gamma}{\Delta\phi\,R}=0,
\end{equation}
with $\gamma$ given by Eq.~\eqref{eq:gamma_tanh}. Because $\mu(r)$ increases
inward, requiring a positive rather than zero value of $\mu$ at the front moves
it to \emph{smaller} $R$; far above threshold the shift is
\begin{equation}\label{eq:deltaR}
	\delta R\simeq-\frac{\gamma}{\Delta\phi\,J_0a}.
\end{equation}

Having obtained the corrected value of $R$, we can now construct the profile valid uniformly 
across the annulus by matching the inner
solution \eqref{eq:domain_wall} to the large-scale solution \eqref{eq:large_scale}:
\begin{equation}\label{eq:phi_an}
\phi_{\rm an}(r)=\phi_\ls(r)-\phi_b\tanh\big[\kappa\,(r-R)\big]-\phi_b\,{\rm sgn}(R-r),
\end{equation}
where the last term removes the common limit $\phi_\ls(R^{\mp})=\pm\phi_b$ of the
two. The same $R$ must be used for the branch switch in $\phi_\ls$ and for the
center of the tanh; otherwise the composite carries a spurious jump of
$2\phi_b$ between the two radii.
\begin{figure}[htbp]
	\centering
	\includegraphics[width=0.48\textwidth]{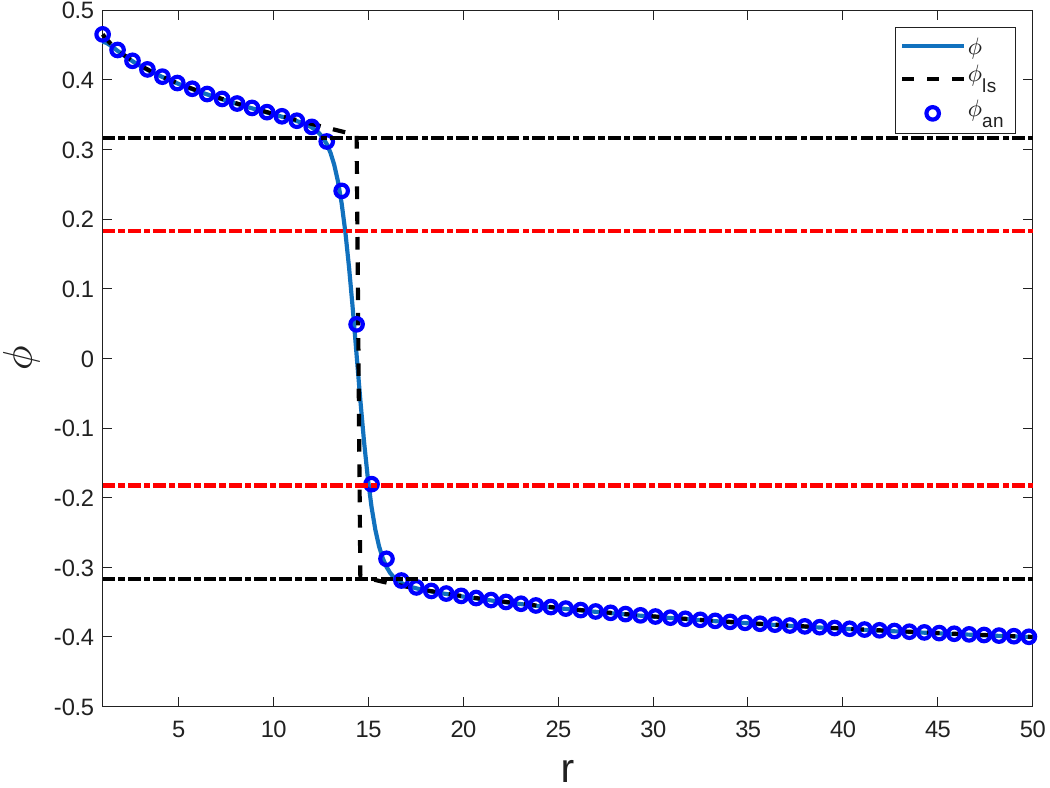}
	\caption{Composition profiles obtained by solving Eq.~\eqref{eq:phi_3rd_order}
		including the square-gradient term. The full blue line is the numerical solution, the
		dashed black curve is the large-scale solution, and the composite
		approximation, Eq.~\eqref{eq:phi_an}, centered on the curvature-corrected
		radius of Eq.~\eqref{eq:R_GT}, appears as symbols. As $\veps\to0$, the
		interface location converges to the sharp-interface radius $R$ of
		Eq.~\eqref{eq:radius_analytical}. The horizontal lines are the bulk spinodal 
		($f_b''(\phi,b)=0$, red) and binodal ($f_b'(\phi,b)=0$, black) values at the given temperature.  
		We used $J_0a=0.02$ and other parameters as in 
		Fig.~\ref{fig_phi_ls_profiles_var_J0}.}
	\label{fig_phi_an}
\end{figure}

Figure~\ref{fig_phi_ls_profiles_var_J0} depicts the algebraic solutions for
$B=0$, where the phase transition manifests as a mathematical discontinuity.
Figure~\ref{fig_phi_an} shows the numerical integration of the full
differential equation with finite $\veps$. The analytical interface (circles) lies extremely close to the 
numerical one (solid line). For the
parameters of Fig.~\ref{fig_phi_an}, the numerical front lies at $R=14.40$,
compared with $15.06$ from Eq.~\eqref{eq:radius_analytical} and $14.52$ from the
curvature-corrected criterion, Eq.~\eqref{eq:R_GT}; the composite analytical solution, 
Eq.~\eqref{eq:phi_an}, deviates from the numerical profile by at most $0.046$, or
7\% of the composition jump $2\phi_b$.

\subsection{Numerical solutions}
\label{sec:numerics}

\begin{figure}[htbp]
\centering
 \includegraphics[width=0.48\textwidth]{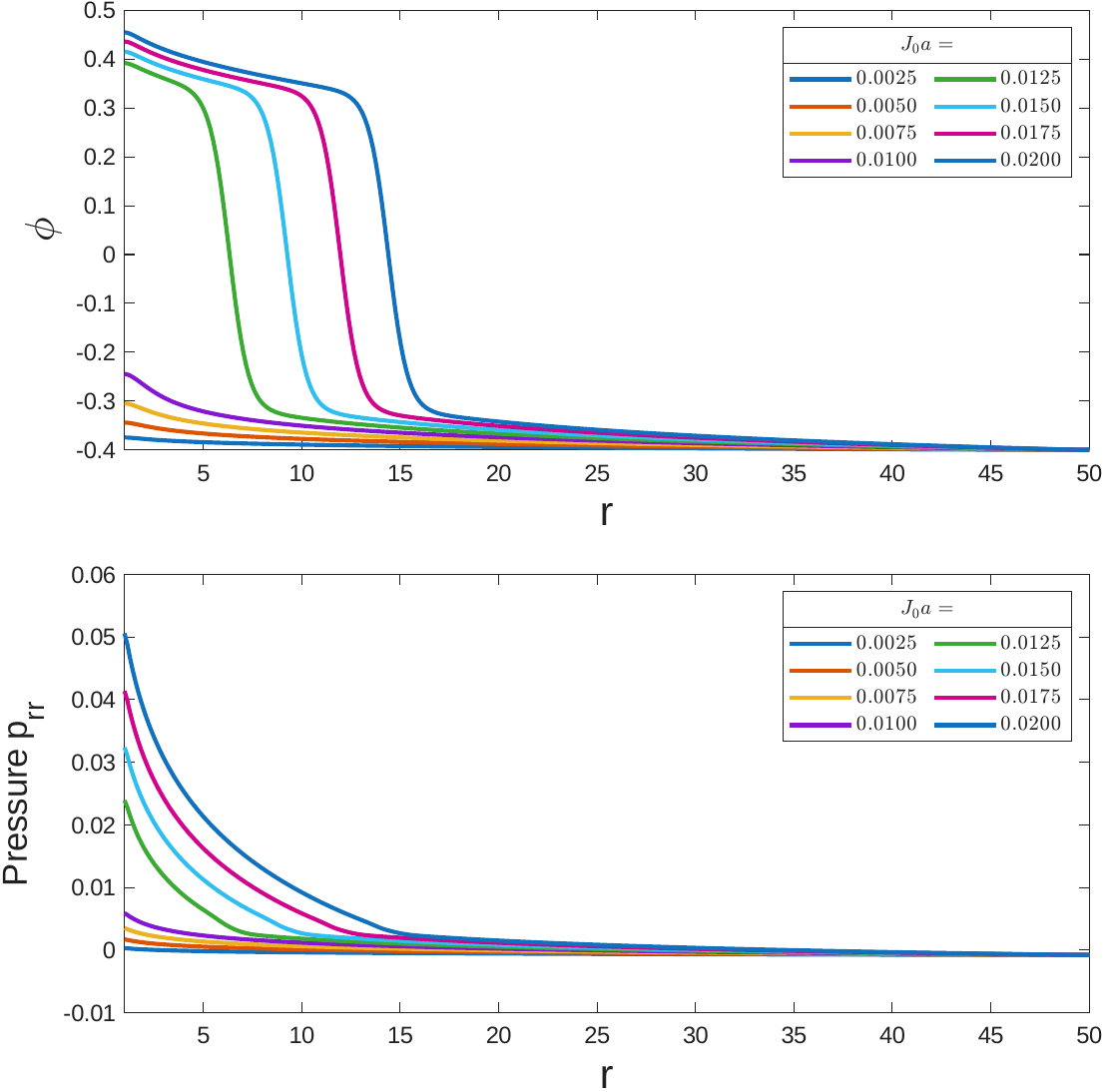}
\caption{(a) Composition profiles obtained by numerical integration of
Eq.~\eqref{eq:phi_3rd_order} for increasing flux. (b) The corresponding radial
pressure $p_{rr}$ from Eq.~\eqref{eq:prr}.
We used the same parameter values as in Fig.~\ref{fig_phi_ls_profiles_var_J0}.}
\label{fig_phi_p_var_J0}
\end{figure}

The steady profiles are obtained by solving Eq.~\eqref{eq:phi_3rd_order} as a
two-point boundary-value problem with the three conditions stated there, using
a collocation solver (MATLAB \texttt{bvp5c}) on a uniform grid of $350$ points
spanning $1\le r\le L=50$. The solution at each flux is used as the initial
guess for the next. With $\veps=0.05$ and $b=-0.1$, the interfacial width is
$\kappa^{-1}=\sqrt{2\veps/(-b)}=1.0$, resolved by about seven grid points.

Figure~\ref{fig_phi_p_var_J0}(a) shows the emergence of a sharp liquid--vapor
interface as the flux is increased. At small flux, the profile is a smooth,
monotonically increasing gas density with no interface. At larger flux, a
condensed annulus at $r<R$ is separated from the vapor at $r>R$ by a wall of
width $\kappa^{-1}$; as the flux grows further, the front moves outward, and the
composition at the inner wall continues to rise above the binodal value because
the local chemical potential keeps increasing.

The sequence in Fig.~\ref{fig_phi_p_var_J0} was computed by stepping
\emph{upward} in $J_0a$ from a front-free initial state, and this matters. The
profiles for $J_0a=0.0075$ and $0.0100$ show no interface, although both exceed
the sharp-interface threshold $(J_0a)_{\rm min}=0.0061$ and the second exceeds
the curvature-corrected threshold of Sec.~\ref{sec:nucleation} as well. Along
an upward sweep, the solver follows the front-free branch, which remains a
solution while it is metastable, and an interface appears only once that branch
ceases to exist. A downward sweep from a condensed state instead retains the
front to $J_0a\simeq0.010$. This is the hysteresis discussed in
Secs.~\ref{sec:nucleation} and~\ref{sec:metastab_hyst}, shown explicitly in
Fig.~\ref{fig_R_var_J0}(b): both branches end at saddle-node bifurcations, and
the two ends bound the hysteresis loop. These ends are definite properties of
the deterministic steady-state problem. They are not, however, the fluxes at
which a real system condenses or evaporates: the calculation contains no
fluctuations, so it cannot nucleate the stable phase, and in practice the
transition occurs somewhere inside the loop.

Figure~\ref{fig_phi_p_var_J0}(b) shows the corresponding $p_{rr}$. Outside the front, the pressure 
is essentially the uniform bulk value; inside, it rises steeply toward $r=1$,
following Eq.~\eqref{eq:mech_balance}. At $r\approx R$, $p_{rr}$ departs
from its plateau value.

\subsection{Shifted binodal lines}
\label{sec:shifted_binodal}

The forced transport renormalizes the local thermodynamic landscape and
enlarges the region of the phase diagram in which a transition can occur.
Condensation first appears at the inner boundary. One can read Eq.~\eqref{eq:J0a_min} as setting the 
temperature for condensation at a given flux $J_0a$. Solving for the temperature variable, we obtain
\begin{equation}\label{eq:bJ}
b_J(\phi)=-\frac{J_0a\ln L}{\phi}-c\phi^2 ,
\end{equation}
and therefore, relative to the equilibrium binodal $b_{\rm coex}=-c\phi^2$,
\begin{equation}\label{eq:bJ_shift}
b_J(\phi)-b_{\rm coex}=-\frac{J_0a\ln L}{\phi}\ >\ 0
\qquad (\phi<0).
\end{equation}
The lines $b_J(\phi)$ lie \emph{above} the binodal: condensation is induced at
temperatures higher, and vapor compositions smaller, than the equilibrium
coexistence. The shift grows linearly with $J_0a$ and
logarithmically with system size.

\subsection{Critical nucleus and the fold}
\label{sec:nucleation}

Equation~\eqref{eq:R_GT} is different from Eq.~\eqref{eq:R_criterion} because $\mathcal{F}$
has a single maximum at
\begin{equation}\label{eq:Rstar}
	R_*=\frac{\gamma}{\Delta\phi\,J_0a},
\end{equation}
and therefore possesses either two roots or none. When two exist, the larger is
the stable front $R$ and the smaller, $R_c<R_*$, is the radius of the
\emph{critical nucleus}: the unstable stationary state separating the
homogeneous gas from the condensed annulus. Two thresholds follow.

\paragraph{(i) Threshold flux.} The two roots merge when
$\mathcal{F}(R_*)=0$, a saddle-node bifurcation. With
$\gamma/(\Delta\phi R_*)=J_0a$, this condition reads
\begin{equation}\label{eq:fold_condition}
	J_0a\left[\ln\!\left(\frac{L\,\Delta\phi\,J_0a}{\gamma}\right)-1\right]
	=-f_b'(\phi_0),
\end{equation}
which can be solved in closed form,
\begin{equation}\label{eq:J0a_fold}
	(J_0a)_{\rm th}=-\frac{f_b'(\phi_0)}{W(z)},
	\qquad
	z=\frac{-f_b'(\phi_0)\,L\,\Delta\phi}{e\,\gamma},
\end{equation}
with $W$ the Lambert function. This is the threshold flux for condensation in the presence of 
surface tension. Comparison with the sharp-interface result, Eq.~\eqref{eq:J0a_min}, shows that
$W(z)$ takes the place of $\ln L$. To see that $W(z)<\ln L$, so that surface
tension raises the threshold, note that $W(z)=-f_b'(\phi_0)/J_0a$ at the fold,
so Eq.~\eqref{eq:fold_condition} reads $\ln(L/R_*)-1=W(z)$, i.e.,
\begin{equation}\label{eq:Rstar_fold}
R_*=L\,e^{-1-W(z)} .
\end{equation}
The fold lies at a physical radius, $R_*\ge1$, only if $W(z)\le\ln L-1$, which
implies $W(z)<\ln L$. If instead $W(z)>\ln L-1$, the fold would lie inside the
bore; $\mathcal{F}$ then decreases monotonically on $1\le R\le L$, the critical
nucleus is always smaller than the bore, and the threshold is given by
Eq.~\eqref{eq:J0a_min_GT} below.

\paragraph{(ii) Barrier-free condensation.} When the critical nucleus is
smaller than the inner hole, $R_c<1$, the wall itself is a supercritical seed: a
liquid film that forms at the wall grows without a curvature barrier. We set $R_c=1$ in
Eq.~\eqref{eq:R_GT} and get
\begin{equation}\label{eq:J0a_min_GT}
	(J_0a)_{\rm sp}^{\gamma}
	=\frac{1}{\ln L}\left[\frac{\gamma}{\Delta\phi}-f_b'(\phi_0)\right].
\end{equation}
This is therefore the \emph{spontaneous} threshold of the curvature mechanism,
not the existence threshold; reading it as ``the flux at which the front reaches the wall'' is
incorrect once curvature is retained.

\begin{figure}[htbp]
	\centering
	\includegraphics[width=0.48\textwidth]{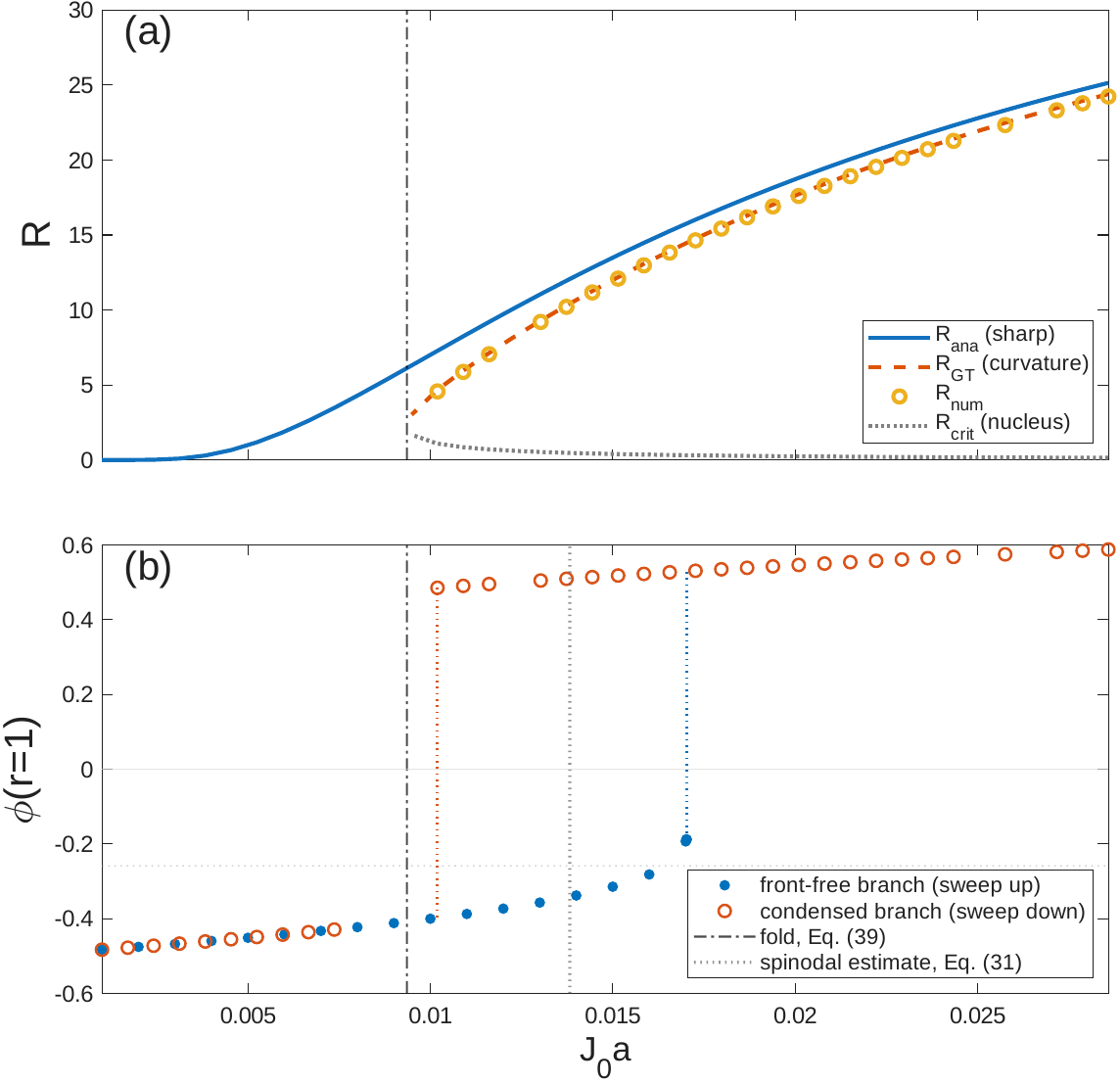}
	\caption{(a) Front location versus flux. Circles: $\phi(R)=0$ from the
		numerical solutions of Eq.~\eqref{eq:phi_3rd_order}, obtained in a downward
		sweep of $J_0a$. Solid line: the sharp-interface prediction,
		Eq.~\eqref{eq:radius_analytical}. Dashed: the stable (large) root of
		Eq.~\eqref{eq:R_GT}. Dotted: the critical-nucleus radius $R_c$, the small,
		unstable root. The vertical dash-dotted line marks the fold,
		Eq.~\eqref{eq:J0a_fold}, where the two roots merge.
		(b) Hysteresis. The wall composition $\phi(r=1)$ for a downward sweep
		(open circles), which follows the condensed branch until it ends at the
		fold, and for an upward sweep from a front-free state (filled circles),
		which follows the front-free branch until it ends at $J_0a\simeq0.0170$.
		Dotted vertical segments mark the jumps between the branches. The vertical
		dotted line is the sharp-kink spinodal estimate, Eq.~\eqref{eq:J0a_sp}, and
		the horizontal dotted line the spinodal composition $-\phi_s$.
		We used $\phi_0=-0.49$, $b=-0.2$, and other parameters as in
		Fig.~\ref{fig_phi_ls_profiles_var_J0}.
	}
	\label{fig_R_var_J0}
\end{figure}

Figure~\ref{fig_R_var_J0}(a) compares the front
location obtained numerically from Eq.~\eqref{eq:phi_3rd_order}, in a downward
sweep of $J_0a$, with the analytical predictions. The numerical front always
lies inside the sharp-interface radius of Eq.~\eqref{eq:radius_analytical}, by an
amount that grows as the flux decreases, in agreement with
Eq.~\eqref{eq:deltaR}. The stable root of the curvature-corrected criterion
in Eq.~\eqref{eq:R_GT} passes through the numerical points over the whole range. The
threshold is reproduced as well: the sharp-interface estimate,
Eq.~\eqref{eq:J0a_min}, gives $0.0050$, roughly half the observed value, whereas
the fold, Eq.~\eqref{eq:J0a_fold}, gives $0.0094$, and the last numerical front
appears at $J_0a\simeq0.0095$. The agreement is better than might be expected,
since near the fold $R\simeq2$ is only about three wall widths and the
sharp-interface expansion is marginal there.

\subsection{Metastability and hysteresis}
\label{sec:metastab_hyst}

Equations~\eqref{eq:J0a_min} and~\eqref{eq:bJ} describe \emph{binodal} crossing. The vapor at
$r=1$ becomes metastable there, not unstable, and the appearance of the
condensed annulus predicted by Eq.~\eqref{eq:radius_analytical} presupposes that
an interface exists. If the system starts from a homogeneous gas, condensation at this
threshold therefore requires a nucleation event, most plausibly heterogeneous at
the inner wall.

Spontaneous condensation, instead, requires the local state to reach the
spinodal, $\phi_s=-\sqrt{-b/3c}$, at which the gas branch of
Eq.~\eqref{eq:large_scale} ceases to exist. The corresponding chemical potential
is
\begin{equation}\label{eq:mu_sp}
\mu_{\rm sp}=f_b'\!\left(\phi_s\right)
=\frac{2}{3}(-b)\sqrt{\frac{-b}{3c}}\ >\ 0 ,
\end{equation}
giving a second, higher threshold
\begin{equation}\label{eq:J0a_sp}
(J_0a)_{\rm sp}=\frac{\mu_{\rm sp}-f_b'(\phi_0)}{\ln L}
=(J_0a)_{\rm min}+\frac{\mu_{\rm sp}}{\ln L},
\end{equation}
and a second family of curves in the $\phi$--$b$ plane, defined implicitly by
\begin{equation}\label{eq:bJ_spinodal}
J_0a\ln L + b\phi+c\phi^3=\frac{2}{3}(-b)\sqrt{\frac{-b}{3c}} .
\end{equation}
Between the two families, the driven gas is metastable: condensation is
thermodynamically favored but kinetically blocked. The transition is then
hysteretic, since the flux required to nucleate the annulus exceeds the flux
below which an existing annulus disappears. 

\begin{figure}[htbp]
\centering
 \includegraphics[width=0.48\textwidth]{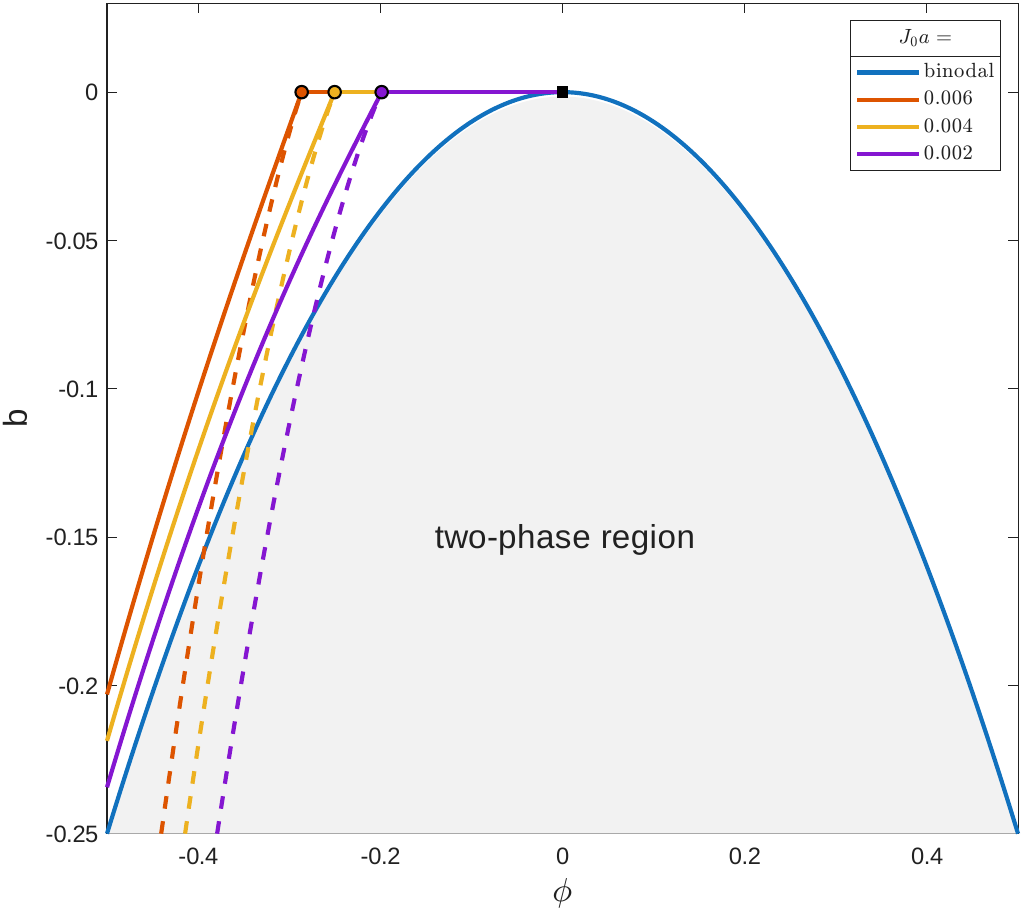}
\caption{Stability diagram in the $\phi$--$b$ plane, with the critical point at
$(\phi,b)=(0,0)$. Blue: the equilibrium binodal $b=-c\phi^2$, below which the
mixture separates with no driving (shaded). The other curves are the
condensation lines, Eq.~\eqref{eq:bJ}, in the presence of a flux. Below these curves, 
liquid condenses at small radii, gas occupies large radii, and an interface
exists at a finite radius $R$; above them, the gas remains single phase, though
with a nonuniform density profile. Dashed: the spinodal family,
Eq.~\eqref{eq:bJ_spinodal}; between a solid curve and the dashed curve of the
same color, the gas at the inner boundary is metastable
(Sec.~\ref{sec:metastab_hyst}). Horizontal segments at $b=0$: the second-order
lines joining each surface critical point (circles) to the bulk critical point
(square). We used $c=1$, $L=50$, and $J_0a=0.002$, $0.004$, and $0.006$.
}
\label{fig_phase_diagram}
\end{figure}

The spinodal family appears as dashed curves in Fig.~\ref{fig_phase_diagram}.
Both families terminate at the same point on the $b=0$ axis, since binodal and
spinodal merge at $T_c$. For the small fluxes shown, much of each dashed curve
lies inside the equilibrium two-phase region, so the metastable band outside
that region is narrow; it widens as $J_0a$ increases.

Interfacial tension produces a second metastable window. When
$(J_0a)_{\rm th}<J_0a<(J_0a)_{\rm sp}^{\gamma}$, a stable annulus exists but is
separated from the homogeneous gas by a barrier, since $1<R_c<R$
(Sec.~\ref{sec:nucleation}), and the transition is again hysteretic. This window
is the curvature counterpart of the bulk binodal/spinodal pair. The two
mechanisms are distinct: the bulk one arises from the loss of local convexity of
$f_b$, the curvature one from interfacial energy, and both operate.

Equation~\eqref{eq:J0a_sp} is a sharp-kink estimate of the upper edge of the
metastable range, and the full solution shows that it is a lower bound.
For the parameters of Fig.~\ref{fig_R_var_J0}, Eq.~\eqref{eq:J0a_sp} gives
$(J_0a)_{\rm sp}=0.0138$, whereas the front-free branch computed in an upward
sweep persists to $J_0a=0.0170$ [Fig.~\ref{fig_R_var_J0}(b)]. The difference
comes from the inner boundary layer: the condition $\phi'(1)=0$ keeps the wall
composition below its large-scale value, so the wall reaches the end of the
front-free branch only at larger flux. At that end, $\phi(1)\simeq-0.19$, beyond
the spinodal composition $-\phi_s=-0.258$; the thin near-wall region that has
crossed the spinodal is stabilized by the gradient term. That the branch ends at
a genuine saddle-node, and not through a failure of the numerics, is confirmed
by the approach of $\phi(1)$ to its end value as
$(J_{\rm end}-J_0a)^{1/2}$.

For the parameters of Fig.~\ref{fig_R_var_J0}, the thresholds are thus ordered
as $(J_0a)_{\rm min}=0.0050<(J_0a)_{\rm th}=0.0094<(J_0a)_{\rm sp}^{\gamma}=0.0104
<(J_0a)_{\rm sp}=0.0138<0.0170$, where the last value is the end of the
front-free branch. The hysteresis loop extends from the fold to that end, and
both barrier-free estimates lie inside it. Above $(J_0a)_{\rm sp}^{\gamma}$, a
liquid film at the wall would grow, but forming it still requires the wall
composition to leave the gas branch, which the bulk barrier prevents until that
branch ends.

\section{Discussion and outlook}
\label{sec:discussion}

We have shown that, in a binary mixture that is stable everywhere at rest, a
steady flux of the order parameter through the inner boundary of an annulus
drives the system out of equilibrium and creates a condensed region bounded by a
stationary front. With diffusion only and no flow,
the steady
state is an equilibrium profile in an effective external field that is
logarithmic in the distance from the center, and the analysis of
Sec.~\ref{sec:caseA} follows from this mapping. 

The logarithmic form of the bias is a property of the geometry rather than of
the constitutive laws. In steady state, conservation alone fixes $rj_r$, and the
bias inherits the logarithm of the two-dimensional Green's function
(Sec.~\ref{sec:general}). In a planar slab, the same argument gives a bias that
is linear in the distance, and the problem reduces to that of a fluid in
gravity~\cite{hohenberg_pra_1972,moldover_rmp1979}. In the classification of
Ref.~\cite{kroll_tsori_jcp2022}, in which the external term in the free energy
has the form $h(\phi)q(r)$, the present case has $h\propto\phi$ and
$q\propto\ln(L/r)$. In the other examples considered there, the field decays
away from its source, whereas here its amplitude at the inner boundary grows as
$\ln L$, so that the outer boundary cannot be sent to infinity. As a result, the
threshold flux, Eq.~\eqref{eq:J0a_min}, vanishes only as $1/\ln L$, and the front
radius, Eq.~\eqref{eq:radius_analytical}, depends exponentially on the ratio of
the bulk driving force to the flux.

The same geometric argument applies to viscous flow through a porous medium. If
the radial flow obeys Darcy's law, $u_r=-(k/\eta)\,dP/dr=-\alpha/r$, the
pressure is logarithmic as well,
\begin{equation}\label{eq:darcy}
P(r)=P_e+\frac{\eta\alpha}{k}\ln\!\left(\frac{r}{L}\right),
\qquad \alpha=\frac{Q}{2\pi w},
\end{equation}
and a first-order transition that is crossed on lowering the pressure occurs at
$R=L\exp[-2\pi wk\,(P_e-P_{\rm tr})/\eta Q]$, where $P_{\rm tr}$ is the transition
pressure. This expression has the same form as Eq.~\eqref{eq:radius_analytical},
with $P_e-P_{\rm tr}$ in place of $-f_b'(\phi_0)$ and $\eta Q/2\pi wk$ in place of
$J_0a$. Near production wells in gas--condensate reservoirs, the falling pressure
produces a bank of condensate that reduces the productivity of the
well~\cite{sayed_spe2016}. This is retrograde condensation, in which liquid
appears when the pressure is lowered. The correspondence with the present model
is structural only: decompression in a liquid mixture is outside of the current model.

The field changes how the interface displaces. In ordinary phase
separation, translating an interface costs no free energy, so domains coarsen
and no length is selected~\cite{bray_advphys1994}. Here, the chemical potential
departs from its coexistence value on either side of the front, so a
displacement of the front costs free energy, and the front is pinned at a
uniquely selected radius (Sec.~\ref{sec_coex}). Arrested coarsening also occurs
in active emulsions, where chemical reactions suppress Ostwald
ripening~\cite{zwicker_pre2015}, and condensates there can be positioned when
sources and sinks maintain a concentration gradient of a third, regulating
component~\cite{weber_rpp2019}. In the present system, the same role is played
by the transport of the phase-separating component itself, with neither a
reaction nor a third component.

The comparison with superfluid helium under heat flow is instructive. There, a
heat current sustains a stationary interface between the normal and the
superfluid phases~\cite{onuki_jltp1983,onuki_jpcm1998}, but it also depresses the
transition temperature~\cite{duncan_ahlers_prl1988} and makes the onset of the
two-phase state first order, even though the transition itself is
continuous~\cite{onuki_jltp1983}. A drive in the bulk can likewise shift the
critical temperature: in a driven lattice gas, it rises with the driving
field~\cite{katz_lebowitz_spohn_jsp1984}. In the present case, the flux changes neither
the critical temperature nor the order of the transition. The effective field
couples linearly to the order parameter, so that the second-order line lies at
the bulk critical temperature and ends at a surface critical point whose
composition, but not whose temperature, is shifted by the flux
(Sec.~\ref{sec:shifted_binodal}. A shift of the critical temperature would
require a coupling that is nonlinear in the order parameter, as for dielectric
mixtures in nonuniform electric fields~\cite{kroll_tsori_jcp2022}.

The case of advection through a rejecting boundary, which describes filtration
and drying, differs in an essential respect. The diffusive flux must then
balance an advective delivery proportional to the local composition
(Sec.~\ref{sec:general}), so the bias depends on the composition itself, and the
steady state is no longer an equilibrium profile in a fixed external field. In
that case, the position of the front must be selected by a criterion that does not rely on
a free energy, for example the existence of a stationary interface between two
homogeneous states, as for flow-induced coexistence in complex
fluids~\cite{olmsted_pre1997}. In membrane science, the critical flux has
already been interpreted as a transition to a condensed phase, with hysteresis
when the flux is raised and then lowered~\cite{bacchin_jms2006}. In models of
colloidal filtration, this transition is placed at the maximum of the osmotic
pressure of the dispersion, that is, at its
spinodal~\cite{aimar_chem_eng_sci_2002}. In the present diffusive problem, by
contrast, the spinodal bounds the hysteresis loop only from above, while the
fold, shifted from the binodal by curvature, sets its lower edge. In drying
films, a dense skin or a demixing front forms at the evaporating
surface~\cite{clark_epl_2007,bouyer_jms_2021}. Whether two saddle-nodes bound an
analogous hysteresis loop in the advective case will be addressed in a
future study.

Interfacial tension generates the nucleation structure of the problem without
further assumptions. The curvature-corrected criterion, Eq.~\eqref{eq:R_GT}, has
two roots, a stable front and a critical nucleus, which merge at a fold,
Eq.~\eqref{eq:J0a_fold}. Below the fold, no condensed annulus exists. Above the
barrier-free threshold, Eq.~\eqref{eq:J0a_min_GT}, the critical nucleus is
smaller than the inner boundary, which then seeds condensation without a curvature
barrier. A similar structure arises around chemically active cores in active
emulsions: a core that produces droplet material guarantees nucleation above a
critical activity, which increases with the surface
tension~\cite{zwicker_pre2015}. There, the source is a reaction, whereas here it
is a flux across the boundary.

The fold and the end of the front-free branch bound a hysteresis loop, which,
for the parameters of Fig.~\ref{fig_R_var_J0}(b), spans nearly a factor of two
in flux. The loop suggests a direct experimental test.
In an annulus whose inner wall is in contact with a permeate of controlled
chemical potential, condensation and evaporation should occur at measurably
different permeate compositions when the permeate is first enriched and then
depleted. Because the calculation contains no fluctuations, the two ends of the
loop are limits of metastability; in a real system, condensation will occur
inside the loop, at a flux set by the kinetics of nucleation.

The model omits the hydrodynamic consequences of the driving. In the steady
state, the stress tensor is not divergence free, Eq.~\eqref{eq:mech_balance}, so
the flux exerts a body force on the fluid that Model B leaves unbalanced. In the
applications discussed above, the condensed phase also impedes the transport that
created it: a condensate bank reduces the productivity of a
well~\cite{sayed_spe2016}, and a deposit on a membrane reduces the permeate
flux~\cite{bacchin_jms2006}. Inclusion of this feedback requires the use of composition-dependent 
mobility $M$, or coupling the composition to a Darcy or a Stokes equation.

Two further questions are left for future work. The first is the stability of
the circular front against azimuthal undulations. The pinning field resists
displacements of the front, but a coupling to flow, together with a viscosity
contrast between the phases, may produce a fingering instability. The second regards a 
non-circular inner boundary. Since the chemical potential is harmonic outside
the front, the front around such a boundary is, at leading order, a level
set of the corresponding harmonic function, and we expect the threshold to be governed by the
logarithmic capacity of the cross section.

\section*{Acknowledgments}

Rudi Podgornik was a great scientist. I heard about him and met him years ago, while I was still a 
young student. Over the years, we had discussions on various scientific topics. His knowledge was 
as wide as it was deep. His concise style often clarified seemingly complex issues. Rudi was living 
proof that one can do excellent, rigorous science while remaining kind, generous, and modest.
His support for us, even in times of real danger, will not be forgotten.
I am grateful for financial support from the Israel Science Foundation Grant No. 332/24, and thank 
H. Diamant for insightful discussions and Juny Tsori for many endearing and enjoyable moments. 

%
%
%

\bibliography{radial_inflow}

\end{document}